\documentclass[aps,pra]{revtex4-2}
\usepackage{braket}
\usepackage{graphicx}
\usepackage{color}
\usepackage{amsmath}
\usepackage{here}
\usepackage{mdframed}
\usepackage{graphicx}

\begin{document}

\title{Tight relation between the physical effects of a quantum measurement and the information gained about an observable}

\author{Natsuki Ogo}
\author{Holger F. Hofmann}
\email{hofmann@hiroshima-u.ac.jp}
\affiliation{Graduate School of Advanced Science and Engineering, Hiroshima University, Kagamiyama 1-3-1, Higashi Hiroshima 739-8530, Japan}

\begin{abstract}

The dynamics of quantum measurements defines a precise relation between the information gained about one physical property of a system and the observable changes in another physical property of the same system. Here, we express this relation in terms of the Hilbert space superpositions of the corresponding eigenstates and show how the probability of an observable physical change can be obtained from the Bayesian update of the probabilities associated with the information obtained in the measurement. Our analysis demonstrates that the superposition principle provides the tightest possible expression of the trade-off between information and back action in a quantum measurement. 

\end{abstract}

\maketitle

\section{Introduction}
\label{sec:Introduction}

It is a well established fact that any measurement performed on a quantum system necessarily involves a corresponding back action on that system. The first formulation of back action by Heisenberg used classical concepts to describe the measurement interaction and left the relation to Hilbert space concepts somewhat unclear \cite{Heisenberg}. Arthurs and Kelly later formulated a quantum mechanical model based on the operation of quantum coherent amplifiers \cite{Art65}, but their theory was not sufficiently general to provide a mathematical formulation of uncertainty relations for quantum measurements. In 2003, the problem was addressed by Ozawa, who obtained a tighter bound for the trade-off between measurement errors and disturbance in generalized quantum measurements \cite{Oza03}. Ozawa's approach to quantum measurements also highlighted a fundamental problem inherent in the definition of measurement uncertainties for physical properties. Is it really possible to evaluate the disturbance of a physical property that did not have a well defined value in the initial state? In response to Ozawa's theory, Busch, Lathi and Werner pointed out that Heisenberg's original bound retains its validity when disturbance is defined in terms of observable changes to the statistics of an observable \cite{Bus13}. A similar result is obtained when the trade-off is based on quantum estimation theory \cite{Wat11,Mat21}. All of these controversies illustrate just how difficult it is to describe the trade-off between measurement information about one observable and the experimentally verifiable physical changes to the eigenvalue of another in terms of the dynamics of measurement interactions \cite{Dre14,Bus14,Roz15,Rod19}. 

A different approach to the information-disturbance trade-off emerged from quantum information theory, where the emphasis was placed on the information encoded in quantum states and the disturbance was understood in terms of the changes to the initial quantum state \cite{Fuc96,Ban01,Ari03,Busc06,Fuc08}. However, the changes to a state itself do not seem to capture the actual physical effects of the measurement process unless we distinguish between physical properties and statistical properties described by quantum states. Information theoretic approaches are dependent on the selection of possible input states as information needs to be encoded in the system \cite{Busc08}. Perhaps the focus on information is not sufficient, as it tends to obscure the physics of the measurement process by over-emphasizing arbitrary choices made in the formulation of specific protocols rather than identifying a specific physical process related to the dynamics of the measurement. A possible compromise is the description of disturbance as decoherence \cite{Sha18}, based on the introduction of a preferred basis by the measurement process itself. It can indeed be shown that the trade-off between the resolution of eigenvalues in a measurement and the associated decoherence between the eigenstate components provides a tight bound for all measurements relating to a specific observable, revealing a fundamental aspect of the relation between information and dynamics in quantum measurements \cite{Pat19}. However, the precise physical meaning of decoherence is itself somewhat elusive, since it has very different effects on different input states. As pointed out by Liu and DeBrota, the relation between different states in Hilbert space should be taken into account when studying the trade-off between information and back action \cite{Liu21}.

The present paper is based on the realization that a gap appears to exist between the original notion of back action as a physical change of the system and the notion of a quantum state as a representation of the statistics of observables. The information-theoretic approach makes it difficult to distinguish between physical changes to the actual properties of a system and Bayesian updates of probabilities associated with the availability of new information. In order to apply Heisenberg's notion of back action to the Hilbert space representation of quantum states, care must be taken to distinguish dynamical effects from the information gained in the measurement. Quantum states make this distinction difficult, because they cannot describe the physics of the system in terms of precise values of its properties. If the state is an eigenstate of one property, the randomness of another physical property will be defined by the expansion in the eigenstate basis of the other physical property. This introduces a precise relation between knowledge of one property and the randomness of another. It needs to be emphasized just how strange this relation between the knowledge of $\hat{B}$ represented by an eigenstate $\ket{b}$ and the probabilities of outcomes $P(a)=|\braket{a|b}|^2$ really is. How can it be that knowledge of one property entails precise knowledge of the {\it probabilities} of another? In the following, we show that we can take this relation as a starting point for a detailed analysis of the observable physical changes to the eigenvalues of a known property $\hat{B}$ required by a Bayesian update of the probabilities $P(a)$ in a quantum measurement. We can then derive the tightest possible relation between the information about a physical property $\hat{A}$ obtained in a measurement and the physical change of the eigenvalue of a property $\hat{B}$ caused by the back action of this measurement. Our results show that the trade-off between information gain and back action in a quantum measurement ultimately originates from the deterministic relation between the precise knowledge of one observable and the necessary randomness of another, revealing how the Hilbert space formalism modifies the relation between dynamics and statistics in a fundamental manner.

The remainder of the paper is organized as follows. In Section \ref{sec:Info}, we express the relation between two observables $\hat{A}$ and $\hat{B}$ in terms of the Hilbert space inner products of their eigenstates and formulate a minimal back action measurement of $\hat{A}$ that corresponds to a general Bayesian update of the information about $\hat{A}$ provided by the input state. In Section \ref{sec:BF}, we relate the back action probability for an eigenstate of $\hat{B}$ to the update fidelity of the property $\hat{A}$. It is shown that the update fidelity provides a lower bound for the mutual information of the measurement, confirming its role as a measure of the information about $\hat{A}$ obtained from the measurement outcomes $m$. In section \ref{sec:PE}, we show that our analysis can also be applied to mixed states using purification, where the back action probability represents the probability of an entangled measurement of the system and the reference. Here, purification effectively defines a global property that can change its experimentally observable value as a result of the physical effects associated with the local measurement performed on the system. In section \ref{sec:MD} we introduce the resolution matrix $R(a,a^\prime)$ of the mesasurement as a convenient tool for the calculation of update fidelity and back action probability for a variety of input states. It is pointed out that the coincidences of outputs in two subsequent measurements of $a$ determine the contribution of each element $R(a,a^\prime)$ to the trade-off and a numerical example is given. Section \ref{sec:conclusion} concludes the paper

\section{Information trade-off between different observables}
\label{sec:Info}

According to quantum theory, knowledge of the precise value of a physical property $\hat{B}$ can determine the probability distribution of measurement outcomes obtained for another physical property $\hat{A}$. This relation between otherwise unrelated physical properties seems rather strange when explained in terms of measurement statistics alone, yet it is an essential part of the Hilbert space formalism where the eigenstates $\ket{b}$ of $\hat{B}$ are represented as superpositions of the eigenstates $\ket{a}$ of $\hat{A}$,
\begin{equation}
\label{eq:expand}
    \ket{b} = \sum_a \braket{a|b} \ket{a}.
\end{equation}
From this representation, it follows that, when the outcome $b$ of a measurement of $\hat{B}$ is already known, the probability distribution over the outcomes $a$ of a precise measurement of $\hat{A}$ can only be 
\begin{equation}
    P(a) = |\braket{a|b}|^2.
\end{equation}
We would like to point out that this is the most fundamental formulation of the uncertainty principle. Quantum mechanics necessarily assigns the same probability distribution $P(a)$ to any situation where $b$ is known. Conversely, this means that any change in our ability to predict the outcomes $a$ must reduce our ability to predict the outcome $b$. Since $b$ is initially known, this loss of information is associated with an observable physical change of the system to different outcomes $b_{\mathrm{out}} \neq b$. 

Since no additional information about $a$ can be obtained without the possibility of physical changes to $b$, interaction free measurements of $a$ are impossible. Instead, all measurement interactions that are sensitive to $a$ must necessarily be accompanied by a back action that changes $b$ through the dynamics induced in the system \cite{Mat21}. Here, we define a measurement of $a$ as an operation on the system that produces an outcome $m$ related to $a$ by a conditional probability $P(m|a)$. We also require that $a$ is not changed by the measurement, so that the conditional probability $P(m|a)$ can be used to update the prediction of outcomes $a$ in a precise measurement of $\hat{A}$ performed on the output. It is then possible to describe the measurement in terms of classical statistics, where the probability of an outcome $m$ is given by 
\begin{equation}
    P(m) = \sum_a P(m|a) P(a).
\end{equation}
and the updated probability of $a$ after the measurement is given by
\begin{equation}
\label{eq:update}
    P(a|m) = \frac{P(m|a) P(a)}{P(m)}.
\end{equation}
In the quantum formalism, the effects of the measurement on an initial quantum state $\ket{\psi}$ are described by Kraus operators $\hat{M}_m$ associated with the outcome $m$. The changes described by such Kraus operators are not limited to the statistical update described by Eq.(\ref{eq:update}), but necessarily include dynamical effects that represent physical changes to the system. In general, measurements could even change the property under observation, but such measurements would necessarily render the Bayesian update in Eq.(\ref{eq:update}) inapplicable. Bayesian updates only apply to so-called quantum nondemolition measurements \cite{Bra80}, where the Kraus operators commute with the target observable $\hat{A}$. The measurement back action that describes the changes to the quantum state is expressed by the eigenvalues of the Kraus operator. The minimal back action required for a conditional probability $P(m|a)$ is described by positive and real eigenvalues equal to the square root of each probability. Additional phase factors represent avoidable back action effects that can be compensated by an appropriate unitary operation \cite{Wis95}. As a result, it is possible to narrow down the Kraus operators of a minimal back action measurement to a from defined entirely by the measurement statistics,
\begin{equation}
\label{eq:Mm}
    \hat{M}_m=\sum_a\sqrt{P(m|a)}\ket{a}\bra{a}.
\end{equation}
These Kraus operators describe the minimal back action required for a measurement of $\hat{A}$ characterized by the conditional probabilities $P(m|a)$. It is not possible to obtain the information needed for a Bayesian update of the probabilities $P(a)$ without the changes to the quantum state described by these operators. All of the observable effects associated with these changes are therefore unavoidable and provide tight bounds on the effects of measurement back action.

The information about $a$ obtained in the measurement results in a Bayesian update of the probabilities $P(a)$ described by (\ref{eq:update}). In terms of the quantum formalism, the measurement probabilities are given by
\begin{equation}
    P(m) = \bra{\psi} \hat{M}_m^\dagger \hat{M}_m \ket{\psi}.
\end{equation}
and the state after the measurement is given by 
\begin{equation}
\label{eq:change}
    \ket{\psi_{\mathrm{out}}(m)}=\frac{1}{\sqrt{P(m)}}\hat{M}_m\ket{\psi}.
\end{equation}
The change of the quantum state reproduces the Bayesian update of Eq.(\ref{eq:update}) when a precise measurement of $a$ is performed after the measurement of $m$,
\begin{equation}
    P(a|m) = |\braket{a|\psi_{\mathrm{out}}}|^2.
\end{equation}
The Bayesian update of probabilities leaves $a$ unchanged, as evidenced by the fact that the initial probability $P(a)$ is recovered from
\begin{equation}
    \sum_m P(m) |\braket{a|\psi_{\mathrm{out}}}|^2 = P(a).
\end{equation}
There is no reason to associate a statistical update with a physical change of the system. To demonstrate that the change of the quantum state described by Eq.(\ref{eq:change}) includes such physical changes, we need to consider a different physical property. 

For any pure state $\ket{\psi}=\ket{b}$, there exists an observable $\hat{B}$ such that the state $\ket{\psi}=\ket{b}$ is a non-degenerate eigenstate of $\hat{B}$. We can only be certain that a measurement of $\hat{B}$ will result in $b$ if the system is in the state $\ket{b}$. Any change of the state $\ket{b}$ can be characterized in terms of the reduction in the probability $P(b_{\mathrm{out}}=b)$ observed in a direct measurement of $\hat{B}$. In a formal sense, this is a way to measure the quantum state fidelity of an arbitrary pure state. However, the measurement procedure is relevant, since it represents the only direct observation of the physical change from one eigenvalue of $\hat{B}$ to another. For a specific measurement outcome $m$, the probability that $b$ is physically changed  by the measurement interaction is
\begin{equation}
    P(b_{\mathrm{out}}\neq b|m) = 1-P(b_{\mathrm{out}}=b|m).
\end{equation}
We will refer to this probability as the back action probability. It should be noted that, different from the changes to the probabilities of $a$, this change cannot be explained by information alone. Specifically,
\begin{equation}
    \sum_m P(m) P(b_{\mathrm{out}}\neq b|m) > 0.
\end{equation}
describes outcomes that were not present in the initial state $\ket{b}$. The information trade-off described by Hilbert space inner products $\braket{a|b}$ can thus be separated into a purely information related change of the statistics and an experimentally observable physical change caused by the interaction dynamics. 

\section{Back action probability and update fidelity}
\label{sec:BF}

Let us first consider the probability $P(b_{\mathrm{out}}=b|m)$ that the measurement did not change $b$ after the outcome $m$ was obtained. The quantum mechanical form of this probability is given by
\begin{equation}
    |\braket{b|\psi_{\mathrm{out}}}|^2 = \frac{1}{P(m)} |\bra{b}\hat{M}_m \ket{b}|^2.
\end{equation}
The spectral expansion of the measurement operator combines the amplitudes of the initial state $\ket{b}$ with the amplitudes of the measurement outcome $\bra{b}$ into probabilities $P(a)=|\braket{a|b}|^2$ that appear to relate exclusively to the initial probability $P(a)$ of the quantum state $\ket{b}$,
\begin{equation}
    \bra{b} \hat{M}_m \ket{b} = \sum_a P(a) \sqrt{P(m|a)}.
\end{equation}
It may be worth noting that any additional phase factors in the eigenvalues of the operator $\hat{M}_m$ can only reduce the probability of finding the unchanged outcome $b$ because the absolute value of a sum of complex numbers with different phases is always smaller than the sum of their absolute values. This observation confirms that Eq.(\ref{eq:Mm}) describes the minimal back action required for a measurement characterized by the conditional probability $P(m|a)$ as originally explained in \cite{Wis95}, even though any additional phase factors would not change the statistical update of $a$.
We find that the probability that $b$ remains unchanged is given by
\begin{equation}
\label{eq:mfidelity}
    P(b_{\mathrm{out}}=b|m) = \frac{1}{P(m)} \left(\sum_a P(a) \sqrt{P(m|a)} \right)^2.
\end{equation}
The right hand side of this equation is defined by the joint statistics of $a$ and $m$ that describe the Bayesian update of $P(a)$ associated with $m$. This means that the physical change of $b$ is related to a measure of statistical similarity exclusively defined by the statistics of $\hat{A}$ before and after the measurement.
\begin{equation}
\label{eq:correspondence}
    P(b_{out}=b|m)=F_A(m).
\end{equation}
The measure of statistical similarity $F_A(m)$ is the statistical fidelity of the initial probability distribution $P(a)$ and its Bayesian update $P(a|m)$,
\begin{equation}
    F_A(m)=\left(\sum_a\sqrt{P(a)P(a|m)}\right)^2.
\end{equation}
In the following, we will refer to this fidelity as the update fidelity of the measurement outcome $m$. 


Eq.(\ref{eq:correspondence}) expresses the first important insight gained from our analysis. It expresses an identity between a measure of physical change to the initial value and a measure of statistical similarity related to the Bayesian update due to the information obtained in the measurement. The identity applies because both of them are derived from the same inner product of initial and final states, $\ket{\psi}=\ket{b}$ and $\ket{\psi_{\mathrm{out}}(m)}$. Although both measures take the form of a statistical fidelity, $P(b_{out}=b|m)<1$ necessarily represent physical changes to the system, since the initial information was maximal and no information gain was possible. Oppositely, an update fidelity of $F_A(m)<1$ represents an update of information about the unchanged outcome $a$. It describes the effects of the information about $a$ gained from the outcome $m$. $P(a)$ and $P(m)$ are the marginals of a single joint probability $P(a,m)$, where the information gain of a readout of $m$ can be described by the mutual information $I(A,M)$. The contribution of each outcome $m$ to the mutual information can be expressed by the relative entropies of $P(a)$ and $P(a|m)$, also known as the Kullback-Leibler divergence $D_{KL}$,
\begin{equation}
I(A,M) = \sum_m P(m) D_{KL}(m).
\end{equation}
where
\begin{equation}
\label{eq:DKL}
D_{KL}(m) = \sum_a P(a|m) \ln \left(\frac{P(a|m)}{P(a)}\right).    
\end{equation}
The relative entropy $D_{KL}(m)$ is zero when the update fidelity $F_A(m)$ is one. Both measures quantify the differences between the marginal probability $P(a)$ and the conditional probability $P(a|m)$. They can be related to each other using the inequality
\begin{equation}
  \frac{1}{2} \ln \left(\frac{P(a|m)}{P(a)}\right) \geq 1-\sqrt{\frac{P(a)}{P(a|m)}}.  
\end{equation}
The sum over $a$ in Eq.(\ref{eq:DKL}) can then be related to the square root of the update fidelity, so that
\begin{equation}
D_{KL}(m) \geq 2 \left(1- \sqrt{F_A(m)}\right).
\end{equation}
A linear relation can be obtained by noting that $2 \geq 1+\sqrt{F_A(m)}$. The update fidelity then provides a lower bound of the relative entropy,
\begin{equation}
\label{eq:KLbound}
D_{KL}(m) \geq 1-F_A(m).
\end{equation}
This inequality indicates that it is more difficult to obtain a low update fidelity than it is to achieve a high relative entropy. As a result, the update fidelity is a better representation of the fundamental information-disturbance trade-off imposed by quantum mechanics. Eq.(\ref{eq:KLbound}) shows that the exact relation between back action probability and update fidelity given by Eq. (\ref{eq:correspondence}) idenitifies the specific part of the update that is responsible for physical effects. The relative entropy is not as selective, combining aspects of the update unrelated to the necessary back action with the contribution described by the update fidelity. In general, the information obtained in an individual measurement is difficult to define, since information gained in the measurement might cancel part of the prior information, resulting in a higher uncertainty after the measurement. Therefore, the information obtained in from a single measurement outcome must always be represented by a statistical distance between the prior and the updated probability. Quantum physics clearly singles out the statistical fidelity as the measure of statistical similarity that efficiently relates the dynamic effects of measurement interactions to the information gained in the measurement.

Our initial result is an equality of the probability of physical change and the change to our knowledge of $a$ caused by a Bayesian update for a specific measurement outcome $m$,
\begin{equation}
    P(b_{\mathrm{out}}\neq b|m) = 1-F_A(m). 
\end{equation}
We can apply this relation to the complete measurement by averaging over all possible outcomes $m$, 
\begin{eqnarray}
    P(b_{\mathrm{out}}\neq b) &=& \sum_m P(m) P(b_{\mathrm{out}}\neq b|m),
    \nonumber \\
    F_{AM} &=& \sum_m P(m) F_A(m). 
\end{eqnarray}
The relation between back action probability and total update fidelity is then given by
\begin{equation}
\label{eq:main}
    P(b_{\mathrm{out}}\neq b) = 1-F_{AM}. 
\end{equation}
This relation is the central result of our analysis - the probability $P(b_{\mathrm{out}}\neq b)$ of an observable physical change of the eigenvalue of $\hat{B}$ in the system  is always equal to one minus the total update fidelity $F_{AM}$ whenever the back action of the measurement is minimized. It should be noted that this is a tight bound for general quantum nondemolition measurements, since any phase factors added to the Kraus operators in Eq.(\ref{eq:Mm}) can only increase $P(b_{\mathrm{out}}\neq b)$. The relation applies to all pure state inputs, since it is always possible to find a specific observable $\hat{B}$ such that the pure state is an eigenstate of the observable with a non-degenerate eigenvalue $B_b$. Here, the difference between one and the total update fidelity obtained by averaging the update fidelities of all possible outcomes describes the information gained about the physically unchanged outcome $a$.

As shown by Eq.(\ref{eq:main}), the total update fidelity of the measurement defines the tight bound for the back action probability of any quantum measurement that realizes a Bayesian update of $\hat{A}$. The mutual information evaluates information differently, making it unsuitable for the formulation of a tight bound on the necessary back action of a Bayesian update. The relation between the two measures of information is given by the inequality
\begin{equation}
    I(A,M) \geq 1-F_{AM}.
\end{equation}
The weight given to different contributions to a statistical update in the mutual information does not correspond to the physics of back action. In general,
The changes to the probabilities of $a$ caused by a measurement of $m$ are based on the information that $m$ provides about $a$. Quantum mechanics relates these changes directly to the probability that an observable physical property changes from $b$ to a completely different value associated with $b_{\mathrm{out}}\neq b$. As we have shown, the relation is precise when the information gain is evaluated in terms of the total update fidelity $F_{AM}$.

\section{Purification and entanglement}
\label{sec:PE}

The discussion above is based on pure state inputs because only pure state inputs can be identified by a characteristic observable property. If the input is in a mixed state, the evaluation of disturbance seems to depend on a subjective element, since we cannot exclude the possibility that the missing information might be available somewhere else. In the context of quantum information theory, this problem is commonly addressed using the method known as purification \cite{Busc06,Sch96}. Here, we adopt the same method and assume that the information missing from a mixed state input is available in the form of pure state entanglement between the system and a reference. In the Schmidt basis, the entangled state is given by
\begin{equation}
    \ket{E}=\sum_n\sqrt{\rho_n}\ket{n}_S\ket{n}_R.
\end{equation}
and the mixed state of the system is
\begin{equation}
    \hat{\rho}_S=\sum_n{\rho_n\ket{n}\bra{n}}.
\end{equation}
If we apply our analysis to the entangled state $\ket{E}$, the probability $P(m)$ depends only on the local state $\hat{\rho}_S$ of the system,
\begin{equation}
    P(m) = \mbox{Tr}\left(\hat{\rho}_S \hat{M}^\dagger_m \hat{M}_m\right).
\end{equation}
On the other hand, the back action probability is defined by the global property $b=E$ of the entangled state. The probability that an entangled measurement of system $S$ and reference $R$ results in $b_{\mathrm{out}}=E$ for a measurement outcome $m$ is given by
\begin{eqnarray}
    P(b_{\mathrm{out}} = E|m) &=& \frac{1}{P(m)} \left|\bra{E} \left(\hat{M}_m \otimes \hat{I} \right) \ket{E}\right|^2
\nonumber \\
&=& \frac{1}{P(m)} \left(\sum_a \sqrt{P(m|a)} \bra{a}\hat{\rho}_S \ket{a}\right)^2.
\end{eqnarray}
This result is the same as that obtained for local pure state inputs in Eq.(\ref{eq:mfidelity}). Even though the result now refers to an entangled measurement, the probability can be determined from the local probabilities of $a$ of the mixed state $\hat{\rho}_S$. All of the relations derived from Eq.(\ref{eq:mfidelity}) apply equally to the purification of a mixed state and to any entangled pure states as well, where the statistics of $a$ refer to the local system $S$ and the physical property $b=E$ is defined globally as a collective property of both system $S$ and reference $R$. Purification and entanglement thus highlight the fundamental difference in the physical meaning of the probability $P(b_\mathrm{out}=b|m)$ and the update fidelity $F_A(m)$ in Eq. (\ref{eq:correspondence}) above. It is always possible that the physical effects of a local information gain can only be observed in global measurements of the system and a quantum reference.

\section{Measurement dynamics and deterministic statistics}
\label{sec:MD}

The total update fidelity $F_{AM}$ depends on the specific combination of initial probability $P(a)$ and conditional measurement probabilities $P(m|a)$. Here, the dynamics of the measurement interaction is completely characterized by $P(m|a)$, independent of the input state $\ket{\psi}$. The dependence of the back action probability $P(b_{\mathrm{out}}\neq b)$ on the initial probability $P(a)$ must originate from the constraint that $P(a)=|\braket{a|b}|^2$ imposes on the dynamics of the state $\ket{b}$. To keep the different contributions to the dynamics separate, we now consider the direct dependence of the total update fidelity $F_{AM}$ on $P(a)$ and $P(m|a)$. We obtain
\begin{equation}
\label{eq:FAM}
    F_{AM} = \sum_{m,a,a^\prime} P(a) P(a^\prime) \sqrt{P(m|a)P(m|a^\prime)}. 
\end{equation}
The sum over the different possible measurement outcomes $m$ can be applied directly to the term characterized by the conditional probability $P(m|a)$. The result of this summation is an input state independent characteristic of the measurement process relating two outcome $a$ and $a^\prime$ to each other by comparing their conditional probabilities. 

As shown in \cite{Pat19}, the decoherence between $\ket{a}$ and $\ket{a^\prime}$ in a quantum measurement is lower bounded by the resolution $R(a,a^\prime)$ defined by the squared Hellinger distance of the conditional probabilities $P(m|a)$ and $P(m|a^\prime)$,
\begin{equation}
    R(a,a^\prime)=\frac{1}{2}\sum_m\left(\sqrt{P(m|a)}-\sqrt{P(m|a^\prime)}\right)^2.
\end{equation}
We can apply this resolution matrix to Eq.(\ref{eq:FAM}) to obtain the update fidelity,
\begin{equation}
\label{eq:resolve}
    F_{AM} = 1-\sum_{a,a^\prime} P(a)P(a^\prime) R(a,a^\prime).
\end{equation}
This equation not only permits a quick calculation of update fidelities for different priors $P(a)$, it also provides a strong intuition of the role of the resolution $R(a,a^\prime)$ since the product of the two probabilities $P(a)$ and $P(a^\prime)$ is equal to the probability of obtaining first $a$ and then $a^\prime$ as outcomes in two independent measurements of $\hat{A}$. The update fidelity is reduced by the resolution times the probability of obtaining this specific pair of outcomes in two independent measurements. It is therefore possible to interpret the resolution $R(a,a^\prime)$ as the reduction in update fidelity associated with a coincidence of $a$ and $a^\prime$ in two consecutive measurements.  

We can now apply Eq.(\ref{eq:main}) to determine the back action probability for an arbitrary input state $\ket{b}$, including possible entangled states as discussed in Sec. \ref{sec:PE} above. For a given resolution $R(a,a^\prime)$, the back action probability is completely determined by the statistics of $a$ in the state $\ket{b}$,
\begin{equation}
\label{eq:disturbance}
    P(b_{\mathrm{out}} \neq b) = \sum_{a,a^\prime} P(a)P(a^\prime)R(a,a^\prime).
\end{equation}
It is worth noting that $R(a,a)=0$ for all measurements. A maximally resolved measurement has $R(a,a^\prime)=1$ for all $a\neq a^\prime$. This determines an upper limit for the back action probability given by
\begin{equation}
    P(b_{\mathrm{out}} \neq b) \leq 1- \sum_a (P(a))^2.
\end{equation}
To understand the physics expressed by this bound, we should remember that $P(a)$ is the necessary probability distribution of $a$ that is required to determine $b$. The back action probability $P(b_|\mathrm{out}\neq b)$ is a consequence of the deterministic relation between the probability $P(a)$ and the physical property $b$. The upper bound defines an intrinsic robustness of $b$ against updates of $P(a)$ given by 
\begin{equation}
    I_{A}(b) = \sum_a (P(a))^2.
\end{equation}
This measure of the intrinsic robustness of $b$ against disturbances caused by measurement of $a$ has a straightforward statistical interpretation as the collision probability of the distribution $P(a)$, which is defined as the probability that two independent trials yield the same outcome. Quantum theory links this statistical property of the distribution of $a$ to the dynamics that change the physical property $b$ in a measurement of $a$. If maximal information is obtained in a measurement projecting the system into the eigenstates $\ket{a}$, the probability $P(b_{\mathrm{out}}=b)$ that the property $b$ does not change during the measurement is given by the collision probability of the distribution $P(a)$ that is required by quantum theory whenever $b$ is determined. 

The precise back action probability depends on the resolution matrix $R(a,a^\prime)$ of the measurement. This matrix quantifies the correlation between the measurement outcomes $m$ and the outcomes $a$ of a precise measurement of $\hat{A}$. To give an impression of the wide range of possibilities that this formalism covers, it may be useful to consider a specific example defined in a three dimensional Hilbert space with $a=1,2,3$. For simplicity, we choose a measurement with only two outcomes, $m1$ and $m2$. For $a=1$, the probabilities are $P(m1|1)=49/50$ and $P(m2|1)=1/50$, For $a=2$, the probabilities are $P(m1|2)=1/50$ and $P(m2|2)=49/50$, and for $a=3$, they are $P(m1|3)=P(m2|3)=1/2$. With these conditional probabilities, the non-zero elements of the resolution matrix are $R(1,2)=R(2,1)=0.72$ and $R(1,3)=R(3,1)=R(2,3)=R(3,2)=0.2$. If this measurement is applied to a maximally entangled state $\ket{E_0}$, the probability that the collective physical property $E_0$ is changed by the measurement is obtained from Eq.(\ref{eq:disturbance}) with an equal probability of $P(a)=1/3$ for $a=1,2,3$. The result is $P(b_{\mathrm{out}} \neq E_0)=0.249$. Note that the intrinsic robustness of $I_A(E_0)=1/3$ would allow a back action probability as high as $0.667$. The reduction of this value to $0.249$ is due to the average resolution of $0.373$ in the six off diagonal elements of the resolution matrix. For comparison, we can now consider an entangled state $\ket{E_1}$ with $P(1)=P(2)=1/2$ and $P(3)=0$. In this case, Eq.(\ref{eq:disturbance}) returns a back action probability of  $P(b_{\mathrm{out}} \neq E_1)=0.36$, equal to the product of the resolution $R(1,2)=0.72$ and the upper bound of $1-I_A(E1)=1/2$. Likewise, an entangled state $\ket{E_2}$ with $P(1)=P(3)=1/2$ and $P(2)=0$
has a back action probability of $P(b_{\mathrm{out}} \neq E_2)=0.1$, equal to the product of the resolution $R(1,3)=0.2$ and the upper bound of $1-I_A(E1)=1/2$. In all three cases, the physical property $b=E_i$ is a collective property of system and reference, yet its disturbance is completely determined by the information gained about the system property $\hat{A}$. Of course, the relations would also be valid for local properties $\hat{B}$, where $\ket{b}$ is a local state of the system. What is important is the tight relation between a probability of physical change and the Bayesian update of probabilities regarding the unchanged property $a$.



\section{Conclusion}
\label{sec:conclusion}

The dynamics of quantum measurements is hard to understand because it is not easy to distinguish between dynamical changes of physical properties and Bayesian information updates. Here, we have addressed this important problem by focusing on the observable physical change of the characteristic eigenvalues of the initial state. We can then obtain a tight relation between the probability of an observable change of $\hat{B}$ and the update fidelity of the probability distribution $P(a)$, as shown in Eq. (\ref{eq:main}). This relation links the back action dynamics of the measurement directly to the representations of eigenstates in Hilbert space, highlighting the fact that the Hilbert space formalism itself requires a non-classical relation between the physical changes induced by the dynamics of a measurement and the information about the system transferred to the measurement apparatus. 

Our analysis reveals several important details about the different roles of input, measurement process and output. The measurement process obtains information about a local physical property $\hat{A}$ of the system, and this information can be verified locally in the system output. On the other hand, the evaluation of the back action probability can involve a quantum reference if the input state is entangled. Here, the non-locality arises from the need to define a precise physical property when all local properties fluctuate. Nevertheless the dynamics of the observable change is described by the local effects of the measurement, where the resolution $R(a,a^\prime)$ between possible outcomes of a precise measurement of $\hat{A}$ necessarily contributes to the probability of an observable change from $b$ to a different value of $b_{\mathrm{out}} \neq b$. 

Eq. (\ref{eq:resolve}) shows how the update fidelity of $\hat{A}$ relates the resolution $R(a,a^\prime)$ to the initial randomness of $\hat{A}$ given by $P(a)$. The information gain of a measurement thus depends on the match between the initial uncertainty of $\hat{A}$ and the sensitivity of the measurement to different outcomes $a$. Since $P(a)$ is fully determined by the Hilbert space representation of $\ket{b}$, the different sensitivities of input states $\ket{b}$ to the back action dynamics of measurements obtaining information about $\hat{A}$ are explained by the non-classical relation between the precise definition of one physical property and the necessary probabilities of another. It seems that this relation is at the heart of all mysteries of quantum mechanics - it is impossible to obtain a classical characterization of the statistics of $\hat{A}$ and $\hat{B}$ because the Hilbert space relations between the two describe dynamics that prevent any meaningful joint observation.

\section*{acknowledgment}
This work was supported by ERATO, Japan Science and Technology Agency (JPMJER2402).

\end{document}